\documentclass[%
reprint,twocolumn,jcp,superscriptaddress,showkeys,floatfix
]{revtex4-1}

\usepackage{graphicx}
\usepackage{dcolumn}
\usepackage{bm}
\usepackage{amssymb}
\usepackage{amsmath}
\usepackage{color}
\usepackage[resetlabels]{multibib}
\usepackage{ifthen}

\newcommand{\fs}[1]{{#1}}
\newcommand{\dir}{.}

\newcommand{\etal}{{\em et al.\ }}
\newcommand{\ud}{{\rm d}}
\newcommand{\DPD}{{\mbox{\tiny DPD}}}
\newcommand{\EOF}{{\mbox{\tiny EOF}}}
\newcommand{\wall}{{\mbox{\tiny wall}}}
\newcommand{\ext}{{\mbox{\tiny ext}}}
\newcommand{\el}{{\mbox{\tiny el}}}
\newcommand{\fluid}{{\mbox{\tiny fluid}}}
\newcommand{\LL}{{\mbox{\tiny L}}}
\newcommand{\CC}{{\mbox{\tiny C}}}

\begin{document}

\title{An efficient dissipative particle dynamics-based algorithm for
simulating electrolyte solutions}

\author{Stefan Medina}
\affiliation{Institut f\"ur Physik, Johannes Gutenberg-Universit\"at Mainz, 
D-55099 Mainz, Germany}
\affiliation{Graduate School Materials Science in Mainz, Staudinger Weg 9, 
D-55128 Mainz, Germany}

\author{Jiajia Zhou}
\affiliation{Institut f\"ur Physik, Johannes Gutenberg-Universit\"at Mainz, 
D-55099 Mainz, Germany}

\author{Zhen-Gang Wang}
\affiliation{Division of Chemistry and Chemical Engineering, 
California Institute of Technology, Pasadena, California 91125, USA}

\author{Friederike Schmid} \email{friederike.schmid@uni-mainz.de}
\affiliation{Institut f\"ur Physik, Johannes Gutenberg-Universit\"at Mainz, 
D-55099 Mainz, Germany}




\date{\today}

\begin{abstract}

We propose an efficient simulation algorithm based on the dissipative particle
dynamics (DPD) method for studying electrohydrodynamic phenomena in electrolyte
fluids. The fluid flow is mimicked with DPD particles while the evolution of
the concentration of the ionic species is described using Brownian pseudo
particles.  The method is designed especially for systems with high salt
concentrations, as explicit treatment of the salt ions becomes computationally
expensive.  For illustration, we apply the method to electro-osmotic flow over
patterned, superhydrophobic surfaces.  The results are in good agreement with
recent theoretical predictions.

\end{abstract}

\maketitle

\section{Introduction} 

Electrolyte fluids are ubiquitous in nature and in (bio)technology. Computer
simulation of electrolyte fluids plays an ever increasing role in research on
soft matter and biophysics  \cite{holm2005advanced, slater2009modeling}. Due to
their complexity, soft matter systems are often described by coarse-grained
models \cite{kremer2003computer, muller2002coarse, nielsen2004coarse,
kroger2004simple}; for example, macromolecules are effectively represented by
bead-spring models. The construction of coarse-grained models capable of
treating both the equilibrium and nonequilibrium behaviors of electrolyte
solutions is a particular challenge. The complex interplay between long-ranged
electrostatic interactions, hydrodynamic interactions, and the
convection-diffusion of ionic species cannot be captured by simple models with
short-ranged effective interactions.  A computationally efficient
coarse-grained approach for simulating the dynamics of electrolyte solutions
must effectively incorporate both types of long-range interactions.  In this
paper, we present one such coarse-grained approach.

We start by a brief review of coarse-grained simulation methodologies for the
different constituents in electrohydrodynamic phenomena. There exist optimized
algorithms for evaluating electrostatic interactions \cite{deserno1998mesh,
groot2003electrostatic, gonzalez2006electrostatic, cisneros2014review}, often
based on Ewald techniques such as the particle-particle-particle mesh
(P${}^3$M) method \cite{hockney1988computer}.  Hydrodynamic interactions can be
treated efficiently by using mesoscale fluid models such as dissipative
particle dynamics (DPD), lattice Boltzmann and multiparticle collision
dynamics \cite{smiatek2012review}.  On the other hand, algorithms have
also been developed that solve the electrokinetic equations directly at the
continuum level \cite{kim2006direct} or on a lattice
\cite{capuani2004discrete}.  Alternatively, particle-based simulation methods
have been proposed that treat either the solvent or the small ions at an
implicit level.  In the former case, the effect of the solvent is described by
a long-range hydrodynamic friction tensor acting between the (explicit) ions
\cite{hernandez2007fast,fischer2008salt}. This approach, however, becomes
inapplicable in systems with arbitrary boundaries or at nonzero Reynolds
numbers.  In the latter case, the ions are replaced by forces acting on the
fluid, which are calculated using the Debye-H\"uckel theory or similar 
approximations \cite{duong2008realistic,hickey2010implicit, hickey2012simulations,
joulaian2012electrospinning}.  
A similar approximation was used to derive a Brownian
dynamics model for polyelectrolytes in solution with both implicit solvent and
implicit ions \cite{kekre2010pressure}.  

Implicit-ion schemes have in common that they assume infinitely fast relaxing
ion clouds, so that retardation effects are not incorporated.  However,
relaxation processes in electric double layers have characteristic time scales
ranging from microseconds to several milliseconds, which are close to the
characteristic time scales of typical mesoscale simulations
\cite{zhou2013dynamic}.  Furthermore, ion distributions can be distorted by
flow.  Thus implicit-ion methods are not suited for studying general dynamical
processes.   On the other hand, explicit treatment of all ions becomes
computationally expensive at high ion concentrations. In this paper, we propose
an alternative approach that uses, as a middle ground, a swarm of virtual Brownian
particles (``pseudo ions") that represent the dynamic ion concentration field.
Such a representation does not properly account for short-range correlations
between the ions on length scales shorter than a coarse-grain length scale that
we can choose, but it preserves the long-range charge-charge correlations.  By
combining this pseudo-ion representation with a mesoscale fluid model and a
fast electrostatic solver, we obtain an efficient tool for the mesoscale
simulation of electrohydrodynamic phenomena in equilibrium and non-equilibrium
soft matter systems.

The rest of our paper is organized as follows: The basic algorithm is developed
in the next section. In principle, the algorithm could be used in conjunction
with any coarse-grained particle-based fluid model (e.g., multiparticle
collision dynamics or coarse-grained molecular dynamics).  Here we focus on DPD
\cite{hoogerbrugge1992dpd, espanol1995dpd}, which is one of the
most popular and widely used coarse-grained simulation methods for complex
fluids. In Section \ref{sec:tests} we discuss first some simple applications
and tests of the algorithm. In Section \ref{sec:patterned_eof}, we demonstrate
the power of the approach on the example of a large-scale problem that cannot
easily be treated by fully explicit simulations: The electro-osmotic flow of
electrolytes with high ionic strength past patterned superhydrophobic surfaces.
We summarize and conclude in Section \ref{sec:conclusions}.

\section{The Condiff-DPD algorithm}
\label{sec:condiff}

We develop our algorithm starting from the set of electrokinetic equations for
electrolyte fluids, which are
(i) the Nernst-Planck equation
\begin{equation}
\label{eq:nernst_planck}
\partial_t \rho_c + \nabla(\rho_c \mathbf{v})
= \nabla \mu_c (e Z_c \rho_c \nabla \Phi + k_B T \nabla \rho_c),
\end{equation}
a convection-diffusion equation
for the number density $\rho_c$ of ionic species $c$ with charge $e Z_c$
and mobility $\mu_c$ in a fluid of velocity $\mathbf{v}$ and
subject to the electrostatic potential $\Phi$; 
(ii) the Poisson equation
\begin{equation}
\label{eq:poisson}
\nabla (\epsilon_m \nabla \Phi)
= - \sum_c e Z_c \rho_c - \rho_{\ext}, 
\end{equation}
for the electrostatic potential where $\epsilon_m$ is
the permittivity of the medium and $\rho_{\ext}$ is the density of fixed
external charges;
and (iii) the Navier-Stokes equation
\begin{eqnarray}
\rho_m (\partial_t \mathbf{v} + (\mathbf{v} \cdot \nabla) \mathbf{v})
  &=& - \nabla P + \eta \Delta \mathbf {v} 
  + (\eta/3 + \zeta) \nabla(\nabla \cdot \mathbf{v}) 
\nonumber \\
&& - e \sum_c Z_c \rho_c \nabla \Phi,
\label{eq:navier_stokes}
\end{eqnarray}
where $\rho_m$ is the mass density, $\eta$ is the shear viscosity, $\zeta$ the
bulk viscosity, and $P$ the pressure (excluding the electrostatic contributions
due to ions, which are accounted for in the last term). The basic idea of our
approach is to simulate the convection-diffusion of ions using a relatively
small number of pseudo ions whose stochastic motion reproduces the ion
distribution described by the Nernst-Planck equation.  Thus we couple a neutral
DPD fluid to Langevin equations for the dynamics of the ions. The electrostatic
potential is calculated using the instantaneous distribution of the pseudo ions
smeared out over the mesh size (see below), which generates a force that enters
the equations of motion for the pseudo ions and the DPD particles.

In practice, we proceed as follows. 
\begin{enumerate}

\item
The Navier-Stokes equations
(\ref{eq:navier_stokes}) are solved through the simulation of an ideal DPD
fluid \cite{hoogerbrugge1992dpd, espanol1995dpd} without
conservative interactions, where DPD particles interact only via dissipative
and stochastic forces.  More complex DPD fluid models could be used as well.
Hydrodynamic boundary conditions at surfaces are implemented with a previously
developed tunable-slip boundary force \cite{smiatek2008tunable}, which also
allows to treat spatially varying surface slip \cite{meinhardt2012separation,
zhou2012anisotropic}.  \fs{The equations of motion for DPD particles are
given in Appendix \ref{app:dpd_equations}.}

\item
The convection-diffusion equation
(\ref{eq:nernst_planck}) for ionic species $c$ is converted into a set of
particle-based equations \cite{szymczak2003boundary}, where the concentration
fields are generated by a cohort of pseudo Brownian particles.   The motion of
these pseudo particles is governed by the Langevin equation
\begin{equation}
\label{eq:langevin_1}
{\rm d} \mathbf{r}_i^c  
- \mathbf{v} \: {\rm d} t
- e Z_c \mu_c \nabla \Phi \: {\rm d} t 
= \sqrt{2 k_B T \mu_c} \: {\rm d} \mathbf{W},
\end{equation}
which includes a convective drift term due to the local velocity field
$\mathbf{v}$ obtained from the DPD particles, a drift term due to the electric
field, and a stochastic term describing the thermal diffusion.  Here
$\mathbf{W}$ denotes a Wiener process \citep{honerkamp1993stochastic}.
\fs{More details on the equation of motion for the pseudo-ions are given
in Appendix \ref{app:ion_equations}.}

\item
The Poisson equation (\ref{eq:poisson}) for the electrostatic
potential is solved by a fast-Poisson solver, which is based on the
particle-mesh approach (see \citep{deserno1998mesh} for an overview) and solves
the equation in Fourier space.  

\end{enumerate}

We use the same assignment scheme as in P${}^3$M \citep{hockney1988computer} to
transfer quantities (charge, force) from the particles to the mesh and vice
versa. Technical details are given in Appendix \ref{app:assignment}. The mesh
also serves as communication hub for the coupling between the pseudo ions and
the DPD fluid.  The mesh defines a coarse-graining length for electrostatic
interactions: Correlation effects on length scales shorter than the mesh size
thus cannot be properly accounted for.  In the simulations described below, the
mesh size was chosen to be of the order of the particle size, and the number of
pseudo ions was chosen equal to the number of real ions in the system. However,
for many applications it is acceptable and may prove advantageous to choose
larger mesh sizes and/or reduce the number of pseudo ions in order to speed up the
algorithm.

Combining all three elements, we obtain an efficient scheme to simulate
electrolyte solutions.  Additional solutes, such as polyelectrolytes, can be
introduced easily.  Because of the central role of the convection-diffusion
process (\ref{eq:nernst_planck}), we call our method the ``{Condiff-DPD}''
method.  

\section{Simple applications and tests}
\label{sec:tests}

We first apply our methods to several simple examples as tests and validation.
The simulation units are given in terms of the thermal energy $k_B T$, the
range $\sigma$ of DPD friction interactions, the mass $m$ of DPD particles, the
elementary charge $e$, and the time unit $\tau = \sigma \sqrt{m/k_B T}$. We
\fs{study} bulk systems with periodic boundary conditions in all directions as well
as slit channels with impenetrable walls at $z=\pm (D/2+\sigma)$ and periodic
boundary conditions in the $x$ and $y$ directions.  The walls interact with DPD
particles and (pseudo-)ions {\em via} a WCA (Weeks-Chandler-Andersen
\citep{weeks1971role}) potential with range $1 \sigma$ and energy parameter
$\varepsilon = k_B T$. Hence the width of the accessible volume in $z$
direction is roughly $D$. The slip length is tuned by applying the tunable-slip
boundary interaction \citep{smiatek2008tunable} (see Appendix
\ref{app:sliplength}). The systems are filled with a DPD fluid of density
$\rho_\DPD$ with DPD friction $\gamma_\DPD=5.0 \tau k_B T/\sigma^2$ and DPD cut-off
$\sigma$.  At $\rho_\DPD=3.75 \sigma^{-3}$ (our most common choice), this
results in a fluid viscosity $\eta = 1.38 \pm 0.03 k_B T \tau/\sigma^3$.  

\fs{We consider electrolyte fluids containing monovalent ions (charge $\pm e$)
with mobility $\mu_c = 0.26 \sigma^2/\tau k_B T$. } The Bjerrum length
$l_B=e^2/4 \pi \epsilon_m k_B T$, which is a measure for the strength of the
electrostatic interaction compared to the thermal energy, is set to be
$\sigma$.  \fs{In the slit geometries, the surface charges are distributed
homogeneously on the wall.} Unless stated otherwise, the number of
pseudo-particles in the Condiff-DPD algorithm is chosen equal to the number of
ions and the mesh size is $a = 0.83 \sigma$. \fs{Details on the method used
to solve the Poisson equation in the bulk and in slit geometry are given
in Appendix \ref{app:poisson}.} The dynamical equations are integrated 
with a time step of $\Delta t=0.01\tau$.  

\fs{For comparison, we have also carried out explicit ion simulations of
electrolytes in slit geometry.  Here, ions are modeled by spherical DPD
particles that carry a unit charge $\pm e$ in the center and mutually repel
each other with a WCA potential of range $\sigma$.  Surface charges are
implemented by randomly placing discrete unit charges in the walls. Previous
simulations have shown that in the weak coupling regime, the electrokinetic
flows on such surfaces are identical to those on homogeneously charged surfaces
\cite{smiatek2009eof}. The simulation are done using the open source program 
package ESPResSo \citep{limbach2006espresso,espressomd}. To calculate
electrostatic interactions, we use the P${}^3$M method with 
electrostatic layer correction (ELC) \cite{arnold2002electrostatics} 
with an ELC gap size of $5 \sigma$
}

\subsection{Bulk electrolytes}
\label{sec:bulk}

\begin{figure}
  \includegraphics[width=0.35\textwidth]{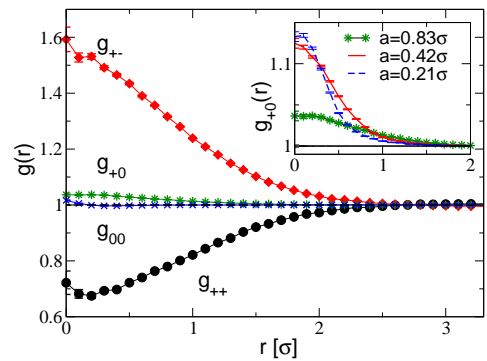}
  \caption{Radial distribution functions 
in a bulk electrolyte with
salt concentration $\rho_i=0.2 \sigma^{-3}$ (all ions)
and DPD fluid density $\rho_\DPD = 3.75 \sigma^{-3}$.
The main frame shows correlations
between positive pseudo-ions ($g_{++}$, black circles),
positive and negative pseudo-ions ($g_{+-}$, red diamonds),
DPD fluid particles ($g_{00}$, blue crosses),
and positive pseudo-ions and DPD particles ($g_{+0}$, green stars)
at grid spacing $a = 0.83 \sigma$. The inset shows 
$g_{+0}$ for different grid spacings.
}
\label{fig:rdf}  
\end{figure}

We begin with studying the structural properties of bulk electrolytes. The
simulations are done in an overall electroneutral system of size $10\sigma
\times 10\sigma \times 10\sigma$ filled with a DPD fluid of density
$\rho_\DPD=3.75 \sigma^{-3}$.  Fig.\ \ref{fig:rdf} shows the pair radial
distribution functions for the various species for a system with total ion
concentration $\rho_i = 0.2 \sigma^{-3}$.  Here the pair distribution is
defined as
\begin{equation}
g_{\alpha \beta}(r) = {\langle \rho_\alpha(r) \rho_\beta(0) \rangle}/
{[\langle \rho_\alpha \rangle \: \langle \rho_\beta \rangle]},
\end{equation}
where \fs{$\alpha,\beta= \pm 1$} represents positive (negative) ions and
\fs{$\alpha,\beta = 0$} represents DPD particles.  The ionic distribution
functions reflect the mutual repulsion of pseudo-ions between like charges (in
$g_{++}(r)$) and the mutual attraction of pseudo-ions between unlike charges
(in $g_{+-}(r)$). The ion-DPD pair distribution $g_{+0}(r)$ features a small
positive correlation: Since the DPD fluid represents both the neutral solvent
and the ions, the DPD density is enhanced in the close vicinity to pseudo-ions;
see discussions in Appendix \ref{app:dpd_profiles}. The range of the
correlation is set by the grid spacing.  With decreasing grid spacing the
correlation becomes steeper and more localized (Fig.\ \ref{fig:rdf}a, inset).
Finally,  inspection of $g_{00}(r)$ shows that DPD particles are basically
uncorrelated, as one would expect for DPD particles \fs{without conservative
interactions}. The indirect correlations induced by the DPD-ion correlations
are too small to be significant. The very small positive correlation close to
$r=0$ \fs{(less than 2 \%)} is an effect of the finite integration time step
which has already been noted in Ref.\ \cite{marsh1997} \fs{and can be removed
by using more sophisticated DPD integrators
\cite{pagonabarraga1998dpd,besold2000dpd,shardlow2003dpd,serrano2006dpd,fabritiis2006dpd}
}.

\subsection{Counterion-induced electro-osmotic flow}
\label{sec:counterion_eof}

Next we study an electrokinetic phenomenon, the electro-osmotic flow
(EOF) \citep{probstein2005physicochemical, lyklema2005fundamentals} in a simple
planar slit geometry. EOF occurs when an external electric field is applied to
an electrolyte solution with net charge, arising, e.g., from the dissociation
of counterions from a surface.  The EOF velocity depends on the slippage at the
surface, which is quantified by the slip length $b$, i.e., the distance between
the hydrodynamic boundary and the point where the flow profile extrapolates to
zero.  In the case of a pure counterion solution both the ion distribution and
the flow profile can be calculated analytically within the Poisson-Boltzmann
approximation.  The theoretical predictions were shown to be in excellent
agreement with previous, fully explicit simulations \cite{smiatek2009eof}.
Hence this problem is a very good test case to validate the new simulation
method.  
 
\begin{figure}
  \includegraphics[width=0.47\textwidth]{\dir/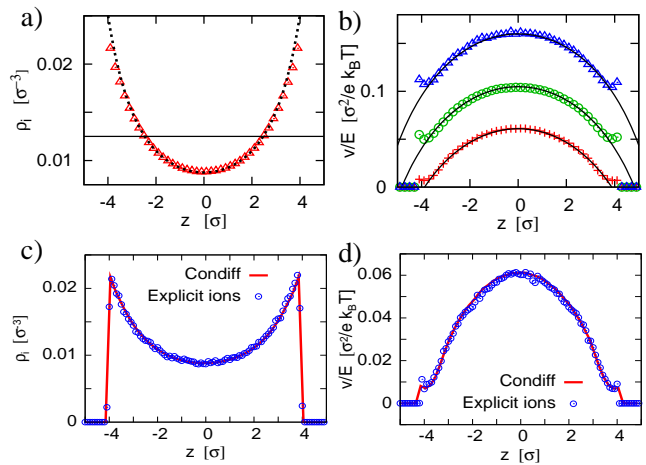}
  \caption{
Counterion-induced EOF in a homogeneous slit channel.  
(a) Counterion distribution according to simulations (red triangles) 
and the Poisson-Boltzmann theory (dotted curve). The density in the center 
$\rho_0$ has been adjusted.  The horizontal line shows the average 
counterion density.  
(b) DPD flow profiles (symbols) and Poisson-Boltzmann result (lines) for slip length 
$b=0 \sigma$ (red crosses), $b=1.21\sigma$ (green circles) and $b=2.65\sigma$ 
(blue triangles).  
c) Counterion distribution according to Condiff-DPD simulations (solid line) and
fully explicit simulations (symbols).
c) DPD flow profile at zero slip obtained from Condiff-DPD simulations (solid line) and
fully explicit simulations (symbols).
}
\label{fig:counterion_eof}  
\end{figure}

In the simulations we consider a slit channel of width $D=8 \sigma$ confined by
charged walls with surface charge density $\Sigma = -0.05 e\sigma^{-2}$ and
filled with the corresponding number of counterions. In order to obtain good
statistics, the simulations shown here are done with 10 pseudo ions per real
ion. We have verified that the results do not depend on the number of pseudo
ions. The area of the simulation box in the $(xy)$ plane is chosen $10\sigma
\times 10 \sigma$. The system is allowed to equilibrate, then flow is induced
by an electric field $E=0.1 k_B T/\sigma e$ in x-direction, and the system is
further equilibrated until steady state is reached. Fig.\
\ref{fig:counterion_eof}a) shows a fit of the counterion profile to the
theoretical Poisson-Boltzmann prediction

\begin{equation}
\rho_i(z) = \frac{\rho_0}{\cos^2(\kappa z)}
\quad \mbox{with} \quad
\kappa = \sqrt{\frac{e^2 \rho_0}{2 \epsilon_m k_B T}},
\end{equation}
where the density $\rho_0$ in the middle of the slit is the only fit parameter.
Knowing $\rho_0$, one can calculate the EOF profiles without further fitting 
{\em via} \cite{smiatek2009eof}
\begin{equation}
\frac{v_x(z)}{E} = -\frac{\epsilon_m k_B T}{Z e \eta}
\left[
\ln\left(\frac{\cos^2(\kappa z_B)}{\cos^2 (\kappa z)}
\right)
- 2 \kappa b \tan(\kappa z_B)
\right].
\end{equation}
The fluid viscosity $\eta$, the slip length $b$, and the hydrodynamic boundary
positions $z_B$, have been determined from independent simulations without
electric field following the procedure described in Ref.\
\citep{smiatek2008tunable}.  We consider three different slip lengths, $b_1=0$,
$b_2=1.21\sigma$ and $b_3=2.65\sigma$. Fig.\ \ref{fig:counterion_eof}b) shows
that the simulation data are in very good agreement with the theory, except in
the close vicinity of the walls, where particles are within the range of
influence of the tunable-slip boundary force \cite{smiatek2008tunable}. 

For comparison, we have also carried out explicit ion simulations of the same
system \fs{with simulation parameters as described above (introduction of Sec.\
\ref{sec:tests})}.  The results are shown in Fig.\ \ref{fig:counterion_eof}c)
and d).  The data obtained with both methods agree very well with each other.

\subsection{Electro-osmotic flow in the presence of salt}
\label{sec:salt_eof}

\begin{figure}
  \includegraphics[width=0.45\textwidth]{\dir/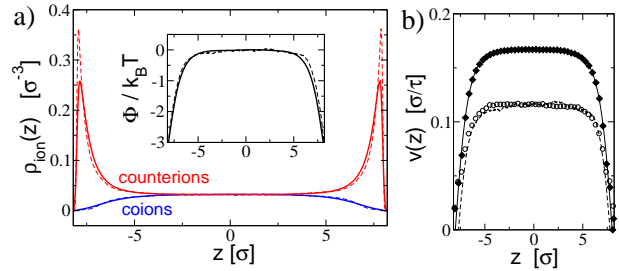}
  \caption{EOF in a homogeneous slit in salt solution with 
 surface charge density $\Sigma = 0.25 e\sigma^{-2}$,
 DPD particle density $\rho_\DPD=3.83 \sigma^{-3}$, and
 ion density (including counterions) 
  $\rho_i = 0.081 \sigma^{-3}$.
(a) Counterion and coion distributions obtained with
the Condiff-DPD algorithm (solid lines) and fully 
explicit simulations with point charges (dashed lines).
Inset shows the corresponding electrostatic potential
as determined from a test particle method.
(b) DPD flow profiles (symbols) compared to the
prediction of Eqn.\  (\ref{eq:theory_eof_vPhi}) (lines). 
Closed symbols/solid lines show data from Condiff-DPD 
simulations, open symbols/dashed lines data from fully
explicit simulations. 
}
\label{fig:salt_eof}  
\end{figure}

Our last test application is EOF in a homogeneous slit 
channel containing salt. For this case, exact analytical solutions of the
Poisson-Boltzmann \fs{equation} are not available. In the Stokes limit,
however, one can still derive an exact relation between the EOF velocity
profile and the electrostatic potential \cite{smiatek2010polyelectrolyte}
\begin{equation}
\label{eq:theory_eof_vPhi}
v_x(z) = \frac{\epsilon_m E_x}{\eta} (\Phi(x)-\Phi_0) + v_\EOF,
\end{equation}
where $\Phi_0$ and $v_\EOF$ are the values of the electrostatic potential 
and the velocity at the center of the slit.
Since $v_\EOF$ depends linearly on the applied electric field $E_x$,
this defines an 'electro-osmotic mobility' 
\begin{equation}
\label{eq:m_eof}
M = v_\EOF/E_x = - \frac{\epsilon_m \Phi_B}{\eta} (1 - \kappa_s b),
\end{equation}
where the 'zeta potential' $\Phi_B$ is the potential at the hydrodynamic
boundary $z_B$ \cite{hunter2000foundations}, and $\kappa_s = \mp \partial_z
\Phi/\Phi \large|_{z_B}$ is the surface screening length \cite{bouzigues2008,
smiatek2010polyelectrolyte}.

Here we consider a slit channel of width $D=16 \sigma$ confined by sticky walls
({\em i.e.}, slip zero) with  surface charge density $\Sigma = 0.25 e
\sigma^{-2}$ and a coion to counterion ratio of 4:9, which results in a total
ion density (all ions) of $\rho_i = 0.081 \sigma^{-3}$.  The area of the
simulation box in $(xy)$ direction was chosen $10 \sigma \times 10 \sigma$.
Fig.\ \ref{fig:salt_eof} shows the ion and velocity profiles, again compared
with fully explicit simulations. Unlike in the example shown in the previous
subsection, the differences between the fully explicit simulations and the
Condiff-DPD simulations are noticeable.  They can be traced back to small
differences in the ion density profiles close to the walls (Fig.\
\ref{fig:salt_eof}a)), \fs{which are most likely caused by the difference of
the ion models. As explained above (introduction of Sec.\ \ref{sec:tests}),
the charges of explicit ions are localized in a point at the center of the ion.}
In the Condiff-DPD model, the charges are smeared out over a range of $\sigma$
(see Appendix \ref{app:poisson}). Therefore, the ion distribution in the
Condiff-DPD simulations is slightly broader and the peak is less high.  This
has a relatively small impact on the electrostatic potential (Fig.\
\ref{fig:salt_eof}a, inset), but a large effect on the EOF (Fig.\
\ref{fig:salt_eof}b). Nevertheless, the shape of the velocity profile $v(z)$
is still in good agreement with the prediction of Eq.\
(\ref{eq:theory_eof_vPhi}) (Fig.\ \ref{fig:salt_eof}b).  We conclude that the
main effect of charge smearing is to renormalize the zeta potential $\Phi_B$.
This must be taken into account when applying the Condiff-DPD algorithm to
systems with high surface charge densities. 

\section{Electro-osmotic flow on patterned superhydrophobic surfaces}
\label{sec:patterned_eof}

\begin{figure}
 \includegraphics[width=0.47\textwidth]{\dir/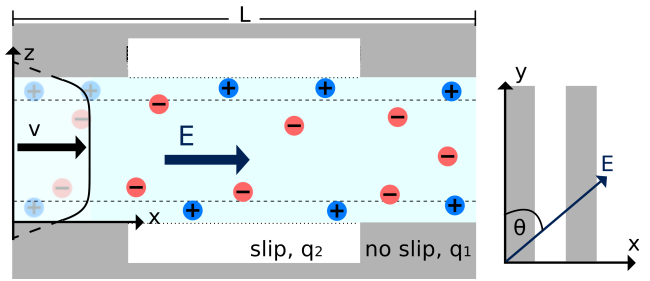} 
 \caption{Left: Cartoon of the a surface with slip/no-slip
pattern and adjacent ion layers, and sketch of the effective plug flow profile in
an electric field (averaged over the whole channel).  Right: top view of
possible flow directions over the striped surface. $\Theta=0$ corresponds to parallel
and $\Theta=\pi/2$ to perpendicular flow.} \label{fig:sketch}
\end{figure}

After these basic tests of the algorithm, we now apply it to an
advanced complex problem: EOF past superhydrophobic surfaces.  Experimentally,
such surfaces contain fractions of gas sectors with a large slip length, which
can be of the order of several microns \citep{ou2005direct}.  The presence of
these gas sectors greatly increases the EOF mobility $M$ and can lead to a
drastic flow enhancement in micron-size channels. If the surface is
anisotropic, the slip length becomes a tensorial quantity
\cite{bazant2008tensorialslip, schoenecker2013} and as a result (as suggested
already by Eq.\ (\ref{eq:m_eof})), $M$ becomes tensorial as well
\citep{belyaev2011electro}.  Here we model superhydrophobic surfaces by plane
walls decorated with a periodic stripe pattern of no-slip and slip areas with
different surface charge as sketched in Fig.\ \ref{fig:sketch}.  The choice of
this particular geometry is motivated by recent theoretical work due to Belyaev
et al \citep{belyaev2011electro}, who provided approximate expressions for the
tensor $M$ in the Stokes limit, using a Debye-H\"uckel approximation and a thin
double layer approximation.  We have studied pressure-driven flow in uncharged
channels with the same geometry in earlier work and verified that the effective
slip reproduces the theoretically expected behavior
\cite{zhou2012anisotropic,asmolov2013}.

In the simulations, we use the same DPD fluid as before in a simulation box of
size $50 \sigma \times 50 \sigma \times 50 \sigma$ with a hard boundary at
positions $z = \pm 25 \sigma$ ({\em i.e.}, the accessible volume is in fact $50
\sigma \times 50 \sigma \times 48 \sigma$), and periodic boundaries in the
other two cartesian directions.  The system is divided into a no-slip sector
(labelled $i=1$) and a slip sector (labelled $i=2$) in the $x$ direction, and
the slip length is varied from no-slip to $b = 15.000 \: \sigma$. This setup
creates a periodic stripe pattern with period $L=50 \sigma$.  Immobile charges
$q_i$ are positioned onto the boundaries in a square grid pattern of spacing
($1\sigma,1\sigma$). The ion (anion and cation) concentrations are chosen such
that the system is overall neutral and the Debye screening length is kept fixed
at $\lambda_D = \sqrt{\epsilon_m k_B T/\sum_c (Z_c e)^2 \rho_c} = 0.99\sigma$
for all systems.  An electric field with amplitude $E = 0.1 k_B T/\sigma e$ is
separately imposed in the x and y directions and the eigenvalues of the
mobility tensor are determined from the velocity of the fluid flow.  The system
is first equilibrated without electric field, the electric field is then turned
on, and data are collected after the system reaches steady-state plug flow.
The ion distributions and flow profiles are results of averaging over $\sim$ 6
million time steps. 

\begin{figure}
\centering
\includegraphics[width=0.47\textwidth]{\dir/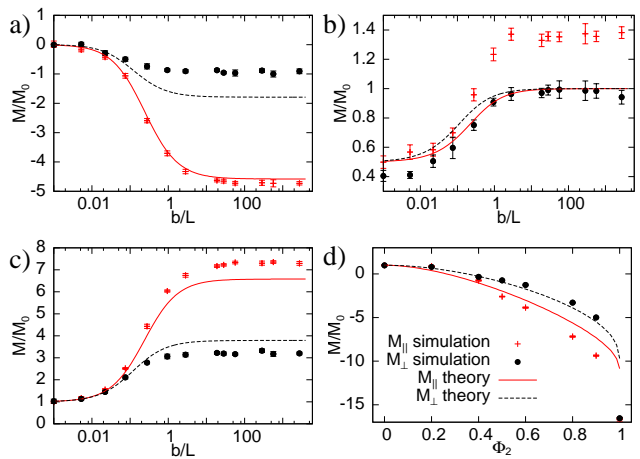}
 \caption{
Normalized EOF mobility on surfaces with geometries as shown in Fig.\
\protect\ref{fig:sketch}. 
The mobility is normalized by the EOF mobility $M_0$ on homogeneous no-slip
surfaces with charge density $q_1$. Curves show the theoretical prediction
(solid for flow parallel to the stripes, dashed for perpendicular flow). 
Symbols correspond to simulation data (red plus for parallel flow, black
circles for perpendicular flow). Frames a-c) show data for stripes of equal width
($\phi_2 = 0.5$) as a function of slip length $b$ in the slip sector for $q_1 =
-q_2 = -0.25 e$ (a), $q_1=-0.5e, q_2 = 0$ (b), and $q_1 = q_2 = -0.25 e$ (c).
Frame (d) shows data for
fixed slip length $b/L = 0.25$ as a function of the slip area fraction $\Phi_2$
for $q_1=-q_2=0.25 e$. The applied electric field is $E=0.1 k_B T/\sigma e$.
} 
\label{fig:eof_patterned} 
\end{figure}

\begin{figure}
\centering
\includegraphics[width=0.47\textwidth]{\dir/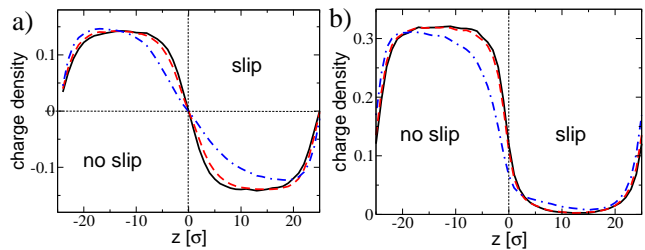}
 \caption{
Net charge density profile adjacent to stripes of equal width with surface
charge distribution $q_1 = -q_2 = -0.25 e$ (a) and $q_1=-0.5e, q_2 = 0$ (b) in
equilibrium systems (black solid lines) and in systems with an electric field
$E=0.1 k_B T/\sigma e$ (red dashed lines) and $E=1.0 k_B T/\sigma e$ (blue
dashed-dotted lines) applied in the direction perpendicular to the stripes.
} \label{fig:ionprofiles} \end{figure}

Fig.\ \ref{fig:eof_patterned} shows representative results for the EOF mobility
parallel and perpendicular to the stripes for different charge distributions
and compares them to the theoretical prediction of Ref.\
\citep{belyaev2011electro} (no fit parameters). Here the mobility data are
normalized by the EOF mobility $M_0$ on homogeneous no-slip surfaces with
charge density $q_1$, which partly removes the uncertainties regarding the
renormalization of the zeta potential $\Phi_B$ discussed in Section
\ref{sec:salt_eof}. In most cases, the data are in very good agreement with
the theory. This is remarkable, given the strong simplifications entering the
theory (Debye-H\"uckel approximation and thin double layer limit). The largest
deviations are found in Fig.\ \ref{fig:salt_eof}b), where the charges are
constrained to the no-slip regions.  Here, the theory significantly
underestimates the amplitude of the parallel EOF mobility for high slip length.
This can be understood when examining the charge distribution in the channel
(Fig.\ \ref{fig:ionprofiles}b).  As expected, a mobile ion layer builds up in
the no-slip region due to the surface charges, and disappears in the slip
region, where the surface charge is zero.  Near the edges, the mobile ions
diffuse into the slip region. The presence of a significant amount of mobile
ions in the slip area leads to an additional contribution to the EOF which is
not taken into account in the theory.

\begin{figure}
\centering
\includegraphics[width=0.47\textwidth]{\dir/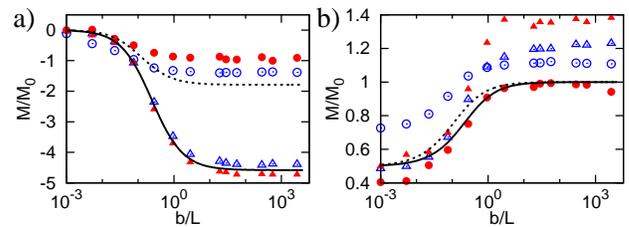}
 \caption{
Same data as in Fig.\ \protect\ref{fig:eof_patterned}a) and b) 
(lines: theory, closed symbols: simulation data for applied field
$E=0.1 k_B T/\sigma e$ ) compared with EOF mobilities at
$E=1.0 k_B T/\sigma e$ (open symbols). Circles refer to perpendicular
flow and triangles to parallel flow. 
} \label{fig:eof_patterned_e1} \end{figure}

The simulation method also allows us to study nonlinear effects in large
electrostatic fields $E$. To this end, we increase $E$ to $E = 1 k_B T/\sigma
e$  in selected cases.  The results for two geometries are shown in Figs.\
\ref{fig:ionprofiles} and \ref{fig:eof_patterned_e1}.  In general, we observe
that nonlinear effects in strong electric fields tend to enhance the
perpendicular EOF mobility $M_\perp$, and reduce the parallel EOF mobility
$M_\parallel$.  This result is somewhat contradictory to a recent theoretical
prediction by Zhao \cite{zhao2010}, who studied systems with charged no-slip
stripes and uncharged full-slip stripes by first order perturbation theory.
According to that study, the leading nonlinear correction should lead to a {\em
reduction} of the perpendicular EOF mobility, due to the fact that the ion
profiles are distorted in the field and this creates an additional restoring
force.  In our simulations $M_\perp$ increases with the electric field; we
suspect that a first-order perturbative treatment is not sufficient at field
strength $E=1.0 k_B T/\sigma e$.  The effect of high field strength on the ion
distribution profiles is shown in Fig.\ \ref{fig:ionprofiles}. They do get
distorted as predicted by Ref. \cite{zhao2010}, but more importantly, the total
amount of counterion charge close to the surface decreases: In strong
perpendicular flow, the mobile ions in the electric double layer are ripped off
the surface and propelled deeper inside the channel, where they can induce EOF
more efficiently.

\section{Discussion and Conclusion}
\label{sec:conclusions}

In summary, we have presented a new simulation approach to studying electrolyte
fluids, which combines a DPD approach for the fluid with a pseudo-particle
representation of the dynamic mean-field equation for the ions. The approach is
designed to optimize the computational efficiency of electrolyte simulations at
high, physiological salt concentrations.  
\fs{Comparing the run times of the Condiff-DPD simulations of 
Sec.\ \ref{sec:patterned_eof} with those of pure DPD simulations with
the same number and density of DPD particles
(but no pseudo-ions, no electrostatics, and no grid),
we find that the time used for electrostatic calculations is less than half
that required for DPD-related calculations, despite the high concentration of
salt ions in the fluid. Hence electrostatic calculations no longer constitute 
the primary bottleneck for large scale simulations of electrolyte fluids.}

The pseudo-particle approach to solving the convection-diffusion equation
matches well with the DPD method and thus creates a fully off-lattice framework
for studying charged fluids. 
\fs{In contrast to other simulation methods that have been designed to solve 
the electrokinetic equation \cite{kim2006direct,capuani2004discrete}, 
the Condiff-DPD method takes a Lagrangian approach that focuses on the 
trajectories of fluid particles and ions. Therefore, it can be combined
naturally with other particle-based simulation models.}
More complex solutes such as macromolecules or
colloids and arbitrary surfaces can be incorporated in a straightforward
manner.  Unlike methods based on the Debye-H\"uckel approximation, our approach
does not assume instantaneous ion-cloud relaxation.

\fs{ Compared to explicit-ion simulations, the Condiff-DPD method has a number of 
advantages: 
\begin{enumerate}
\item
It does not scale with the number of ions. The costs for
electrostatic calculations are dominated by the costs for solving the Poisson
equation and depend primarily on the number $M$ of mesh points (with scaling $M
\log M$). In comparison, the integration of the Langevin equation for
pseudo-ions, Eq.\ (\ref{eq:langevin_1}) or (\ref{eq:langevin_2}), is fast due
to the absence of direct pair interactions between pseudo-ions.
Moreover, the number of pseudo ions can be chosen independently from the number
of real ions \cite{footnote}. Therefore, the Condiff-DPD method is particularly
suited for studies of electrolyte solutions at high concentrations. 
\item
It correctly reproduces the purely
diffusive character of micro-ion dynamics in Eq.\ (\ref{eq:nernst_planck}),
i.e., the fact that the inertia of micro-ions is irrelevant on the time scales
of interest.  (typical relaxation times are picoseconds or less
\cite{Grossmann2012electrophoresis}) -- in contrast to coarse-grained 
explicit-ion simulations, where the finite mass of ions introduces unphysical 
inertia effects \cite{zhou2012colloid}.  
\item 
Possibly as a consequence of 2., we observe in our EOF
simulations that the equilibration time to reach a stationary state
is much shorter in Condiff-DPD simulations than in explicit-ion simulations.
This also significantly speeds up the simulations.
\item 
Ion properties such as the ion mobility are input parameters 
and can be tuned at will -- in contrast to standard molecular dynamics methods, 
where they have to be determined from simulations. 
\end{enumerate}
}
We remark that the Condiff-DPD approach is not restricted to electrolyte
solutions. The same idea can be applied to other mesoscale fluid flow
simulations where the diffusion of a minority component is important, e.g., in
\fs{microreactors}.

\acknowledgments{ The authors thank Olga Vinogradova and Aleksey
Belyaev for providing the theoretical data and Burkhard D\"unweg for
fruitful discussions. This work was funded by the Volkswagen Stiftung.
Simulations with explicit ions were performed using 
the open source program package ESPResSo \citep{limbach2006espresso,espressomd}
Simulations were carried out at the NIC Computing Center J\"ulich and
on the Computer Cluster Mogon at Mainz University.
}

 \begin{appendix}

\fs{
\section{Summary of equations of motion in the Condiff-DPD algorithm}

For the convenience of the reader, we summarize the equations of motion that
were implemented in the algorithm.

\subsection{DPD equations for fluid particles}

\label{app:dpd_equations}

The fluid particles (DPD particles) follow Newton's equation of motion
$m \ddot{\mathbf{r}}_i = \mathbf{F}_i$, where the force $\mathbf{F}_i$
acting on particles $i$ has three contributions:
\begin{equation}
\label{eq:DPD_forces}
\mathbf{F}_i = \mathbf{F}_{\CC,i} + \mathbf{F}_{\DPD,i}
+ \mathbf{F}_{\LL,i} + \mathbf{F}_{\el,i}.
\end{equation}
The first term, $\mathbf{F}_{\CC,i}$, summarizes all conservative forces.
In our simulations, DPD particles have no conservative interactions with
each other. However, in slit geometry, they interact with walls in slit 
{\em via} a repulsive WCA potential \cite{weeks1971role}, 
$\mathbf{F}_{\CC,i} = \mathbf{F}_{\wall}(\mathbf{r}_i)$ with
\begin{equation}
\label{eq:wall_force}
\mathbf{F}_\wall(\mathbf{r}_i)
= 4 \varepsilon \Big(
(\frac{r_0}{d_i})^{12} - (\frac{r_0}{d_i})^6 + \frac{1}{4}
\Big)
\quad \mbox{for $d_i < r_0$,}
\end{equation}
where $d_i$ is the closest distance between $\mathbf{r}_i$ and the wall.
We use $\varepsilon = 1 k_B T$ and $r_0 = 1 \sigma$.

The second term in Eq.\ (\ref{eq:DPD_forces}), $\mathbf{F}_{\DPD,i}$,
corresponds to the DPD thermostat, which is a sum of pairwise forces
with a dissipative and a stochastic contribution 
\cite{hoogerbrugge1992dpd,espanol1995dpd}
\begin{eqnarray}
\label{eq:DPD_thermostat}
\mathbf{F}_{\DPD,i} 
&=& \sum_{j \ne i} 
\big\{
- \gamma_\DPD \: \omega_\DPD^2(r_{ij}) \: (\hat{r}_{ij} \mathbf{v}_{ij}) \:
\hat{r}_{ij}
\\ \nonumber
&& + \: \sqrt{2 \gamma_\DPD k_B T} \: \omega_\DPD(r_{ij})
\hat{r}_{ij} \:\zeta_{ij}(t)
\big\}.
\end{eqnarray}
Here $r_{ij} = |\mathbf{r}_{ij}|$ with
$\mathbf{r}_{ij} = \mathbf{r}_i - \mathbf{r}_j$ is the distance
of particles $i$ and $j$,
$\hat{r}_{ij} = \mathbf{r}_{ij}/r_{ij}$ the unit vector
pointing from $j$ to $i$, and 
$\mathbf{v}_{ij} = \mathbf{v}_i - \mathbf{v}_j$ 
their relative velocity. We use the standard form for the 
DPD weight function, $\omega_\DPD(r) = \omega(r,\sigma)$ with
\begin{equation}
\label{eq:weight}
\omega(r,r_c) = \left\{ 
\begin{array}{ll} 1-r/r_c & : \; r < r_c \\ 0 &: \; r > r_c \end{array}
\right.,
\end{equation} 
The DPD friction parameter $\gamma_\DPD$ tunes the
shear viscosity of the fluid; here we choose 
$\gamma_\DPD=5 \tau k_B T/\sigma^2$.
Finally, $\zeta_{ij}(t)$ describes a Gaussian distributed white
noise with $\langle \zeta_{ij}(t) \rangle = 0$ and
$\langle \zeta_{ij}(t) \zeta_{kl}(t')\rangle = 
 \delta(t-t') \: (\delta_{ik} \delta_{jl} + \delta_{il}\delta_{jk})$.
The DPD thermostat preserves momentum by construction, hence a DPD
fluid follows the Navier Stokes equation \cite{hoogerbrugge1992dpd}.
Furthermore, the dissipative and the stochastic contribution 
satisfy the fluctuation-dissipation relation, hence DPD particles
are Boltzmann distributed at equilibrium \cite{espanol1995dpd}.

The third contribution to Eq.\ (\ref{eq:DPD_forces}), $\mathbf{F}_{\LL,i}$,
describes the friction between fluid particles and the wall and
takes the form \cite{smiatek2008tunable}
\begin{equation}
\label{eq:tunable_slip}
\mathbf{F}_{\LL,i} = 
- \gamma_\LL \omega_\LL^2(d_i) \: \mathbf{v}_i 
+ \sqrt{2 \gamma_\LL k_B T} \: \omega_\LL(d_i) \: \mathbf{s}_i(t),
\end{equation}
where $d_i$ is the closest distance between particle $i$ and the
wall (as in Eq.\ (\ref{eq:wall_force})), the weighting function
is chosen $\omega_\LL^2(d) = \omega(d, 2 \sigma)$ 
(using Eq.\ (\ref{eq:weight})), the vector $\mathbf{s}_i(t)$ is 
a Gaussian distributed white noise with 
$\langle \mathbf{s}_i(t) \rangle = 0$ and
$\langle s_{i\alpha}(t) s_{j \beta}(t') \rangle
= \delta(t-t') \: \delta_{ij} \delta_{\alpha \beta}$
($\alpha, \beta = x,y,z$), and the parameter $\gamma_\LL$
can be used to tune the slip of the fluid at the wall
\cite{smiatek2008tunable} (see Appendix \ref{app:tunableslip}).
Again, the dissipative and stochastic terms in Eq.\ 
(\ref{eq:tunable_slip}) are constructed such that they
satisfy the fluctuation-dissipation relation.

Finally, the last force contribution in Eq.\ (\ref{eq:DPD_forces}), 
$\mathbf{F}_{\el,i}$, describes the effect of electrostatic forces 
on the fluid (the last term in Eq.\ \ref{eq:navier_stokes}). It is
evaluated at the level of the pseudo-ions and transmitted
to the DPD particles following a procedure described in
Appendix \ref{app:coupling}.

In our simulations, the DPD equations were integrated with a 
standard Velocity-Verlet scheme where the stochastic contribution
was included {\em via} a simple Euler algorithm.

\subsection{Langevin equation for pseudo-ions}

\label{app:ion_equations}

Pseudo-ions and DPD particles do not interact directly with
each other. They are coupled exclusively through the grid
as described in Appendix \ref{app:coupling}. The motion of pseudo-ions
in the electric field $\mathbf{E} = -\nabla \Phi$ is described
by the Langevin equation (\ref{eq:langevin_1}), which has
to be supplemented by the contribution of wall forces in 
the case of slit simulations. The resulting equation of motion 
for pseudo-ions representing the ionic species $c$ 
can be written as 
\begin{equation}
\label{eq:langevin_2}
\dot{\mathbf{r}}_i = \mathbf{v}_{\fluid}(\mathbf{r}_i)
+ \mu_c\big(e Z_c \mathbf{E}(\mathbf{r}_i) 
  + \mathbf{F}_{\wall}(\mathbf{r}_i) \big)
+ \sqrt{2 \mu_c k_B T} \: \mathbf{s}'_i(t),
\end{equation}
where $e Z_c$ is the ionic charge, $\mu_c$ the ionic mobility, and the vector
$\mathbf{s}'_i(t)$ is another Gaussian distributed white noise satisfying, as
usual, $\langle \mathbf{s}'_i(t) \rangle = 0$ and $\langle s'_{i\alpha}(t)
s'_{j \beta}(t') \rangle = \delta(t-t') \: \delta_{ij} \delta_{\alpha \beta}$
($\alpha, \beta = x,y,z$). The velocity field $\mathbf{v}_{\fluid}
(\mathbf{r})$ is the local velocity of the DPD fluid which is transmitted to
the pseudo-ion system through the grid as described in Appendix
\ref{app:coupling}. The electric field $\mathbf{E}(\mathbf{r})$ is also
calculated on the grid based on the distribution of pseudo-ions, see Appendix
\ref{app:poisson}. The additional force $\mathbf{F}_{\wall}(\mathbf{r})$ is
taken from Eq.\ (\ref{eq:wall_force}) and accounts for the confinement in slit
simulations. 

Two points must be stressed. First, pseudo-ions don't interact directly with
each other, only through the grid. Hence, the equations of motion can be
integrated very efficiently.  Here we used a simple Euler forward algorithm.
Second, Eq.\ (\ref{eq:langevin_2}) corresponds to overdamped dynamics: In
contrast to DPD particles, pseudo-ions have no mass. This takes into account
that the pseudo-ions are used to model a convection-diffusion equation without
inertia, Eq.\ (\ref{eq:nernst_planck}), whereas the DPD particles model the
Navier-Stokes equations, Eq.\ (\ref{eq:navier_stokes}), where inertia is
included and important.  

}

\section{Transferring quantities between particles and the mesh}

\label{app:grid}

The Condiff-DPD algorithm relies on an efficient communication between 
particles and a grid. Firstly, the grid is used to calculate the
electrostatic forces: Pseudo-ion charges are distributed onto neighboring 
grid points, the Poisson equation is solved on the grid, and the 
resulting electric field is redistributed onto the pseudo ions. 
Secondly, the grid is responsible for coupling the pseudo-ions 
with the the DPD fluid: Electric forces acting on pseudo-ions are 
collected by the grid points and redistributed onto the DPD particles. 
Conversely, velocities of DPD particles are assigned to grid points
and used to determine the local fluid velocity which enters the
equation of motion of the pseudo-ions.

\subsection{Assignment scheme}
\label{app:assignment}

\begin{figure}[tbp]
\centering
\includegraphics[width=0.25\textwidth]{\dir/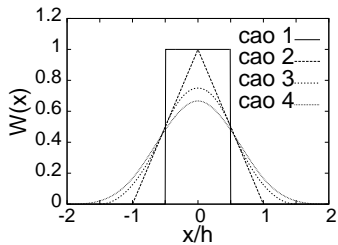}
\caption{Series of cardinal B-splines used as assignment functions. 
Increasing the assignment order (cao) leads to smoother assignments 
but larger support. In the present work we use cao 3.}
\label{fig:cao}
\end{figure}

As mentioned in the main text, we use the same assignment scheme as in
P${}^3$M\cite{hockney1988computer} to transfer quantities between
particles and the grid. We now provide a brief description of the procedure. Given a set of
particles with position $\mathbf{r}_i$ carrying a quantity $Q_i$, the process
of assigning densities to grid points can be described by the sum
\cite{deserno1998mesh}

\begin{equation}
\rho_Q(\mathbf{r}_p) = 
\frac{1}{h^3} \sum_i Q_i W(\mathbf{r}_i-\mathbf{r}_p), 
\end{equation}
(in three dimensions), where $h$ is the grid spacing and $W(\mathbf{r})$ a
weighting function which is chosen such that the quantity $Q$ is always
conserved -- i.e., the fractions of $Q_i$ distributed among the grid points
must sum up to the total value of $Q_i$.  Specifically, we choose a third order
cardinal B-spline assignment scheme \cite{hockney1988computer} (see Fig.\
\ref{fig:cao}), which corresponds in Fourier space to the weight function
$\hat{W}(\mathbf{k}) = \tilde{W}(k_x) \tilde{W}(k_y) \tilde{W}(k_z)$ with

\begin{equation}
\tilde{W}(k) = h \left( \dfrac{\sin(k h/2)}{k h/2}\right)^3.
\end{equation}
The real-space implementation of this weight function is easily done with
piecewise polynomials.  Further details including a table of this
representation up to order seven can be found in Ref.\ \cite{deserno1998mesh}.
The cardinal B-spline assignment corresponds to the original choice by Hockney
and Eastwood \cite{hockney1988computer} in their P${}^3$M method. Other, similar
methods such as PME and SPME use different assignment schemes. It is important
to note that for reasons of consistency, the same assignment scheme has to be
used for transferring particle properties to grid points and vice versa.

\subsection{Solving the Poisson equation on the grid}
\label{app:poisson}

For a given charge distribution $\rho_q$ in infinite space, the solution of the
Poisson equation, $\Delta \Phi = - 4 \pi \frac{l_B k_B T}{e^2} \rho_q$, 
in Fourier space is given by 
\begin{equation}
\label{eq:greensfunc}
\hat{\Phi}(\mathbf{k}) = \frac{l_B k_B T}{e^2}  
\hat{G}(\mathbf{k}) \: \hat{\rho}_q(\mathbf{k})
\end{equation}
with the Green's function $\hat{G}(\mathbf{k}) = 4 \pi/k^2$.  We take the
charges on pseudo-ions to be smeared out according to a Gaussian distribution.
In Fourier space, the particle shape is thus described by the smearing function
\begin{equation}
\tilde{\gamma}(\bm{k}) = \exp(-k^2/4\delta^2)~~.
\end{equation}
Unless stated otherwise, we choose $\delta = 1/\sqrt{2} \sigma$.
However, charges are assigned to grid points based on the positions
$\mathbf{r}_i$ of the centers of the pseudo ions. This introduces an error. The
error can be minimized by modification of the Green's function in Eq.
(\ref{eq:greensfunc}) to a so-called optimal Green's function $G_\text{opt}$,
which takes into account the assignment scheme, the extension of the charges
and the specific type of differentiation operator used. Luckily, one only
needs to know the k-space shape of $G_\text{opt}$, which according to
Hockney and Eastwood is given by
\begin{equation}
\tilde{G}_\text{opt}(\bm{k}) = \dfrac{ \bm{\tilde{D}}(\bm{k}) \cdot
\sum_{\bm{m} \in \mathbb{Z}^3} \tilde{U}^2\left(\bm{k} +
\frac{2\pi}{h}\bm{m}\right) \bm{\tilde{R}}\left(\bm{k} +
\frac{2\pi}{h}\bm{m}\right) } { \left| \bm{\tilde{D}}(\bm{k}) \right|^2 \left[
\sum_{\bm{m} \in \mathbb{Z}^3} \tilde{U}^2\left(\bm{k} +
\frac{2\pi}{h}\bm{m}\right) \right]^2 }~~.
\end{equation}
Here $\bm{\tilde{D}}(\bm{k})$ is the k-space version of the differentiation
operator used in the algorithm, $\tilde{U}=\hat{W}(k)/h^3$ is the k-space
version of the assignment function divided by the volume of one cell, and
$\bm{\tilde{R}}(\bm{k})$ is the true reference force acting on the smeared
charges that is to be modeled with the grid approach, i.e., 
\begin{equation}
\label{eq:referenceforce}
\bm{\hat{R}}(\bm{k}) = - i \bm{k} \hat{G}(\bm{k})\hat{\gamma}^2(\bm{k}),
\end{equation}
with the true Green's function $\hat{G}$ and the differentiation operator
$-i\bm{k}$. This expression must replace the ordinary Green's function to
minimize the error in the force.  (A different shape is obtained if one
minimizes the error in the energy.) We note that Eq.\ (\ref{eq:referenceforce})
differs slightly from the corresponding expression proposed by Deserno and Holm
in Ref.\ \cite{deserno1998mesh}.  The reason is that Deserno and Holm view the
k-space part of P${}^3$M as an interaction between a smeared particle
and a point charge. In the original approach, however, Hockney and Eastwood
take it to be the interaction between two smeared charges, which is precisely
the situation we want to model here.

\fs{ The method described so far solves the Poisson equation in bulk systems
with full periodic boundary conditions.  In simulations of slit channels, the
systems under consideration are non-periodic in one direction (the $z$
direction). To account for this, we use the ELC approach
\cite{arnold2002electrostatics}, which starts from the P${}^3$M solution for
the slab system with periodic images in $z$ direction separated by a finite
(empty) gap, and then systematically substracts the contributions of these
periodic images. The ELC method is formally exact. Nevertheless, we found that
it may produce uncontrollable numerical errors in systems with high local
surface charges, most notably in the system studied in Fig.\
\ref{fig:eof_patterned}b). Therefore, we also implemented an approximate
correction term proposed by Yeh and Berkowitz \cite{yeh1999elc}.  At gap size
$5 \sigma \approx 5 \lambda_D$, simulations using the Yeh-Berkowitz correction
gave the same results as simulations based on the full ELC correction in all
cases where ELC was stable. Furthermore, the Yeh-Berkowitz correction did not
produce numerical instabilities in the system of Fig.\
\ref{fig:eof_patterned}b).  Therefore, the data presented in Sec.\
\ref{sec:patterned_eof} refer to simulations of slab systems separated by gaps
of width $5 \sigma$ with a Yeh-Berkowitz correction term \cite{yeh1999elc}.  }

\subsection{Using the grid to couple the DPD fluid and the pseudo ions}
\label{app:coupling}

Every communication between DPD particles and pseudo ions is mediated by
the grid, i.e., a quantity to be transmitted is transferred to the grid first
and then redistributed to the other species. 

For the fluid velocity this is implemented as follows:
\begin{enumerate}
\item Assign velocity and number density of DPD particles to the grid,
  following the procedure described above.
\item Normalize velocity by dividing by the DPD number density.
\item Distribute normalized velocity from the grid onto pseudo-particles.
\end{enumerate}

For the electrostatic force, the procedure is as follows:
\begin{enumerate}
\item Assign the electric field from the grid to particles.
\item Multiply by pseudo-ion charge to obtain the electric
  force acting on pseudo-ions.
\item Assign electric forces acting on pseudo-ions to the
  grid.
\item Assign number densities of DPD particles to the grid.
\item Normalize forces by the DPD number density.
\item Distribute normalized force field from the grid to
  the DPD particle.
\end{enumerate}
Step 5 is necessary to ensure that the force acting on 
pseudo-particles matches exactly that transferred to the
DPD particles. 

\section{Discretization errors}

\subsection{Accuracy of particle mesh electrostatics}
\label{app:accuracy_ewald}

\begin{figure}[tb]
\centering
\includegraphics[width=0.4\textwidth]{\dir/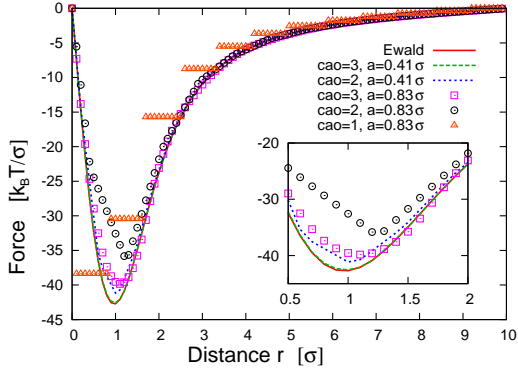}
\caption{
Electrostatic force in a system containing one positive and one negative charge
as calculated from particle-mesh calculations with different assignment orders (cao)
and grid spacings $a$, compared to the result obtained by accurate Ewald summation.
}
\label{fig:ewald_pairforce}
\end{figure}

The accuracy of particle-mesh Ewald techniques depends on the parameters of the
algorithm such as the choice of charge assignment scheme and the grid spacing.
A number of authors have developed sophisticated schemes to estimate the error
of electrostatic force calculations \cite{hockney1988computer,petersen1995accuracy,
deserno1998mesh, deserno1998mesh2}. These schemes were originally designed for
systems of point charges, where interaction potentials have to be split in a
smeared long range part, which is treated in Fourier space, and a residual
short range part, which is treated in real space. In our model, the pseudo-ion
charges are smeared already. This simplifies the analysis considerably, since
the interactions can be treated fully in Fourier space.

To assess the influence of the grid spacing and the charge assignment order on the
accuracy of electrostatic interactions, we consider a bulk system of size $20
\sigma \times 20 \sigma \times 20 \sigma$ containing one positively charged and
one negatively charged particle. Here the charges are smeared with
$\delta^{-1} = \sigma$. We calculate the interactions between the charges for
different charge assignment order (cao, see Fig.\ \ref{fig:cao}) and grid
spacing $a$, and compare the results with direct Ewald sums in Fourier space 
using a very large cutoff in $k$-space. The resulting curves are shown in Figure
\ref{fig:ewald_pairforce}. Except for charge assignment order cao 1, all curves
agree in the long-range limit. Deviations set in around $r < 2 \sigma$, when
the particles overlap. They can be reduced by choosing a smaller grid spacing $a$
and/or a charge assignment of higher order.  At cao 3 and $a=0.83 \sigma$ (the
parameters chosen in most of this work), the particle-mesh calculations 
are in reasonable agreement with the results of the exact Ewald summation.

\subsection{DPD density profiles}
\label{app:dpd_profiles}
 
\begin{figure}[tbp]
\centering
\includegraphics[width=0.4\textwidth]{\dir/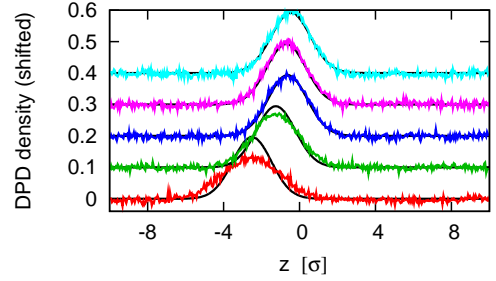}
\caption{
Ion density profiles (black lines) compared with DPD profiles,
$(\rho_\DPD - \rho_s)$ in an electrolyte system subject to an
external electrostatic potential $\Phi(z) \propto z^2$ 
for different grid spacings $a$. Profiles 
for different $a$ values are shifted in $z$ and the $y$ axis
for better visibility. Grid spacings from top to bottom are
$a=0.08 \sigma$,
$a=0.28 \sigma$,
$a=0.42 \sigma$,
$a=0.83 \sigma$,
and $a=1.67 \sigma$.
}
\label{fig:dpdprofiles_pot}
\end{figure}

As already noted in the main text (Section \ref{sec:bulk}), the DPD particles
represent the total density of the fluid, including neutral solvent and small
ions.  To illustrate this, we consider an equilibrium system (stationary and
$\mathbf{v} = 0$) in an electrostatic potential $\Phi$.  Eq.\
(\ref{eq:nernst_planck}) ensures that the ions are Boltzmann distributed,
$\rho_c = \rho_c^0 \exp(-\Phi/k_B T)$. Combining this with Eq.\
(\ref{eq:navier_stokes}), one obtains the equilibrium relation $k_B T \sum_c
\nabla \rho_c - \nabla P = 0$. Since our DPD particles have no conservative
interactions, they have ideal gas statistics and the local pressure (excluding
electrostatics) satisfies $P = \rho_\DPD k_B T$.  Hence the ''neutral solvent
density'' $\rho_s = \rho_\DPD - \sum_c \rho_c$ fulfills $\nabla \rho_s \equiv
0$, {\em i.e.}, $\rho_s$ is constant everywhere as one would expect.  
The shape of the DPD profile therefore follows that of the total ion profile:
\begin{equation}
\label{eq:dpd_distribution}
\rho_\DPD = \rho_s + \sum_c \rho_c
= \rho_s + \sum_c \rho_c^0 \exp(-\Phi/k_B T).
\end{equation}
The distribution $\rho_\DPD$ defines an effective potential acting on DPD particles,
$\Phi_\DPD = - k_B T \ln(\rho_\DPD)$, from which one can derive
an effective force $\mathbf{F}_\DPD = - \nabla \Phi_\DPD$. The latter can be
related to the total force acting on all ions,
$\mathbf{F}_i = - \sum_c \rho_c \nabla \Phi$, {\em via}
\begin{equation}
\label{eq:dpd_force}
\mathbf{F}_\DPD = - \mathbf{F}_i / \rho_\DPD.
\end{equation}
This equation is consistent with the requirement that the total force
acting on the DPD fluid must match the total force acting on all pseudo ions
(Section \ref{app:coupling}).

Ideally, the coupling scheme between pseudo ions and DPD particles should
ensure both an accurate distribution coupling (Eq.\
(\ref{eq:dpd_distribution})) and an accurate force coupling (Eq.\
(\ref{eq:dpd_force})). Unfortunately, this is not feasible for finite grids.
Only one type of coupling can be implemented exactly.  Since momentum
conservation is crucial in hydrodynamics, the Condiff-DPD algorithm was
constructed to ensure exact force coupling.  This automatically leads to
discretization artefacts in the distribution of DPD particles.

To illustrate and investigate these effects, we study
a bulk system of size $10\sigma \times 10 \sigma \times 20 \sigma$
exposed to a quadratic external electrostatic potential $\Phi(z) \sim z^2$.
The ions are then localized in the potential well, and the DPD profile
reflects the peak of the ion distribution. We vary the grid spacing 
between $a=0.08\sigma$ and $a = 1.67 \sigma$. The resulting profiles 
for charge assignment order cao3 are shown in Fig.\ \ref{fig:dpdprofiles_pot}.
At larger grid spacings, the DPD profile does not fully follow the 
ion profile. Reasonable agreement is obtained for $a=0.83 \sigma$,
which is the standard grid spacing used in the present work.

\begin{figure}[tbp]
\centering
\includegraphics[width=0.47\textwidth]{\dir/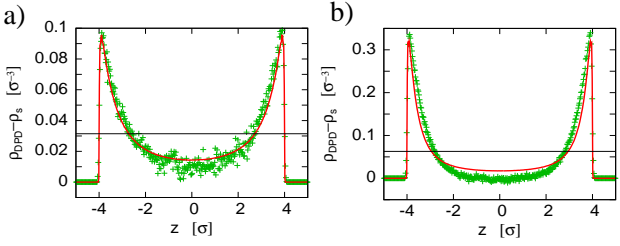}
\caption{
Counterion density profiles (black solid line) and shifted DPD density 
profiles (symbols) in a slit channel confined by charged plates
with surface charge density $\Sigma = 0.125 \sigma^{-2}$ (a)
and $\Sigma = 0.25 \sigma^{-2}$ (b).
}
\label{fig:dpdprofiles_eof}
\end{figure}

The artifacts in the DPD profiles are most pronounced close to impenetrable
walls, where high surface charges may generate large charge gradients. We
find that this leads to an excess of DPD particles close to the surfaces. Fig.\
\ref{fig:dpdprofiles_eof} compares ion profiles and shifted DPD density
profiles in a slit channel of width $D=8 \sigma$ containing counterions only
for different surface charge densities $\Sigma$. At low $\Sigma$, the  profiles
match well (Fig.\ \ref{fig:dpdprofiles_eof}a).  At higher $\Sigma$, a
disproportionately high number of DPD particles accumulates at the walls. For
extreme surface charges $\Sigma > 4/\sigma^2$, the DPD particles may even
withdraw entirely from the bulk of the slab. 
%

This problem can easily be remedied either by using smaller grid spacings or by
introducing repulsive interaction between DPD particles, {\em i.e.}, making the
fluid less compressible. In the applications presented in this paper, however,
this was not necessary because the deviations of the ideal DPD profile shapes
were small. The DPD density profiles are not the focus of interest in this
work. We only need to ensure that the DPD fluid reproduces the correct fluid
dynamics. Thus the algorithm must guarantee a faithful force transmission
between pseudo-ions and DPD particles, and the shear viscosity of the
fluid should be roughly constant. This is still the case, even if the DPD
density varies slightly. 

\section{Tunable slip algorithm applied to electrolytes}

\label{app:tunableslip}

\begin{figure}[t]
\centering
\includegraphics[width=0.47\textwidth]{\dir/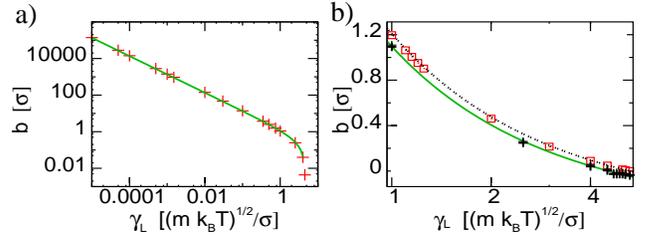}
\caption{
(a) Slip length versus amplitude of boundary friction $\gamma_L$ obtained
from simulations of an electrolyte fluid with Debye screening length
$\lambda_D=0.99 \sigma$ close to a wall with surface charge density $\Sigma =
0.25 \sigma^{-e}$ (symbols), compared to the analytical fit to Eq.\
(\protect\ref{eq:slip_neutral}) with renormalized DPD density
$\overline{\rho_\DPD}$ (line) and fit parameter $C$.
(b) Comparison of the slip length in the same electrolyte fluid
(lower curve, solid line and plus symbols) with that in a
neutral DPD fluid (upper curve, square symbols and dashed line).
Symbols correspond to simulation data, lines are theoretical fits 
to Eq.\ (\protect\ref{eq:slip_neutral}) with fit parameter $C$ and,
(in the electrolyte case) renormalized density 
parameter $\overline{\rho_\DPD}$.
}
\label{fig:sliplength} \end{figure}

The tunable slip algorithm by Smiatek \etal \cite{smiatek2008tunable} is
designed to implement hydrodynamic boundary conditions with arbitrary slip
lengths in DPD simulations. This is done by introducing a thin viscous layer
covering the walls, which is implemented in terms of a locally varying friction
function.  The slip length is controlled by the friction between the fluid and the
wall, more precisely by the shape, the amplitude, and the range of the
friction function.  For example, if one assumes that the fluid density
$\rho_\DPD$ is constant up to a hard boundary at $z=0$ and if one chooses a
linearly varying friction function $\gamma(z) = \gamma_L (1-z/z_c)$, a good
analytical estimate of the slip length is provided by the formula
\cite{smiatek2008tunable}
\begin{equation}
\label{eq:slip_neutral}
\frac{b}{z_c} = \frac{2}{\alpha} + C
- \frac{19}{1800} \alpha
+ \frac{293}{772625} \alpha^2 + \cdots
\end{equation}
with $\alpha = z_c^2 \gamma_L \rho_\DPD/\eta$ and $C = -7/15$. For neutral fluids,
this equation is reasonably accurate at high $b$, but becomes problematic 
at low $b$. Corrections based on numerical simulations have recently
been proposed by Zhou \etal \cite{zhou2013trapezoid}. 

ln our model electrolyte fluids, the density of the DPD fluid increases close
to charged walls (see Appendix \ref{app:dpd_profiles}). Therefore $\rho_\DPD$
is no longer constant, and Eq.\ (\ref{eq:slip_neutral}) no longer applies.
Nevertheless, we find that it can still be used after a simple heuristic
modification: It suffices to replace the density $\rho_\DPD$ in the
dimensionless factor $\alpha$ by an effective density
\begin{equation}
\overline{\rho_\DPD} = \frac{\int_0^\infty \ud z\: \gamma(z) \: \rho_\DPD(z)}
      {\int_0^\infty \ud z\: \gamma(z)}.
\end{equation}
In practice, we fit the function $\rho_\DPD(z)$ by a second order polynomial before
performing the integration. Furthermore, we allow for a uniform offset in Eq.\
(\ref{eq:slip_neutral}), which provides us with a single fit parameter $C$.
With these adjustments, Eq.\ (\ref{eq:slip_neutral}) is found to fit the
simulation data for the slip length over a wide range of $b$ (Fig.
\ref{fig:sliplength}a). Here we have simulated an electrolyte fluid with Debye
screening length $\lambda_D = 0.99 \sigma$ close to a homogeneous charged
surface with surface charge density $\Sigma = 0.25 \sigma^{-2}$.  The slip
length data are obtained by a joint analysis of Poiseuille and Couette flows
following Ref.\ \cite{smiatek2008tunable}. The density renormalization leads to
a slight decrease of $b$ for given friction amplitude $\gamma_L$ (Fig.
\ref{fig:sliplength}b)).

\label{app:sliplength}

\end{appendix}

\bibliography{slabeof}

\begin{thebibliography}{10}
\bibitem{footnote}%
  \BibitemOpen
  \bibinfo {note} {This holds in the Poisson-Boltzmann (the mean field) regime.
  If charge fluctuations are important, the number of pseudo-particle should
  match the number of real ions.}%
  \bibAnnoteFile{Stop}{footnote}%
\end{thebibliography}

\end{document}